\documentclass[aps,prl,superscriptaddress,reprint,notitlepage]{revtex4-2}

\usepackage{bm,color}
\usepackage{graphicx}
\usepackage{amsmath, amssymb}
\usepackage{braket}
\usepackage{comment}
\usepackage[compat=1.1.0]{tikz-feynman}
\usepackage{txfonts}

\usepackage[utf8]{inputenc}

\usepackage[hidelinks]{hyperref}

\usepackage{hyperref}
\usepackage{xcolor}
\hypersetup{colorlinks=true}
\usepackage{graphicx}
\usepackage{bm}
\usepackage{amsmath}
\usepackage{amssymb}
\usepackage{xspace}
\usepackage{algorithmic}
\usepackage{algorithm}
\usepackage[capitalise]{cleveref}
\usepackage{txfonts}
\graphicspath{{./figures/}}
\usepackage{physics}
\usepackage{siunitx}
\usepackage{booktabs}
\usepackage{chemformula}
\usepackage{ulem}
\usepackage{orcidlink}
\newcommand{\subfigref}[2]{Fig.~\hyperref[#1]{\ref*{#1}#2}}

\begin{document}
\title{
    Rhombohedral Multilayer Graphene as a $p$-Wave Magnet
}
\author{Rintaro Eto\orcidlink{https://orcid.org/0000-0001-8833-1311}}
\email{rintaro.eto@tum.de}
\affiliation{Technical University of Munich, TUM School of Natural Sciences, Physics Department, 85748 Garching, Germany}
\affiliation{Munich Center for Quantum Science and Technology (MCQST), Schellingstraße 4, 80799 München, Germany}
\author{Johannes Knolle\orcidlink{https://orcid.org/0000-0002-0956-2419}}
\affiliation{Technical University of Munich, TUM School of Natural Sciences, Physics Department, 85748 Garching, Germany}
\affiliation{Munich Center for Quantum Science and Technology (MCQST), Schellingstraße 4, 80799 München, Germany}
\date{\today} 
\begin{abstract}
    We propose rhombohedral multilayer graphene in an applied in-plane magnetic field as a highly tunable platform for nonrelativistic collinear $p$-wave magnetism. The orbital coupling to the magnetic field breaks time-reversal symmetry and, together with interaction-driven layer antiferromagnetism, generates an odd-in-momentum spin splitting without relying on spin-orbit coupling. Using a minimal low-energy effective theory, we show that the resulting $p$-wave spin splitting is strongly enhanced with increasing layer number. This enhancement originates from the surface-localized nature of the low-energy states and their layer-dependent orbital coupling to the in-plane field. The Zeeman coupling only weakly perturbs this predominantly orbital-field-induced mechanism. We further estimate stability of the $p$-wave magnetism under hole-doping and distinct transport signature using a more realistic Hubbard model. Our results establish orbital coupling to an external magnetic field as a controllable route to odd-parity spin splitting in collinear magnets.
\end{abstract}
\maketitle

\begin{figure}[hbt]
    \centering
    \includegraphics[width=0.48\textwidth]{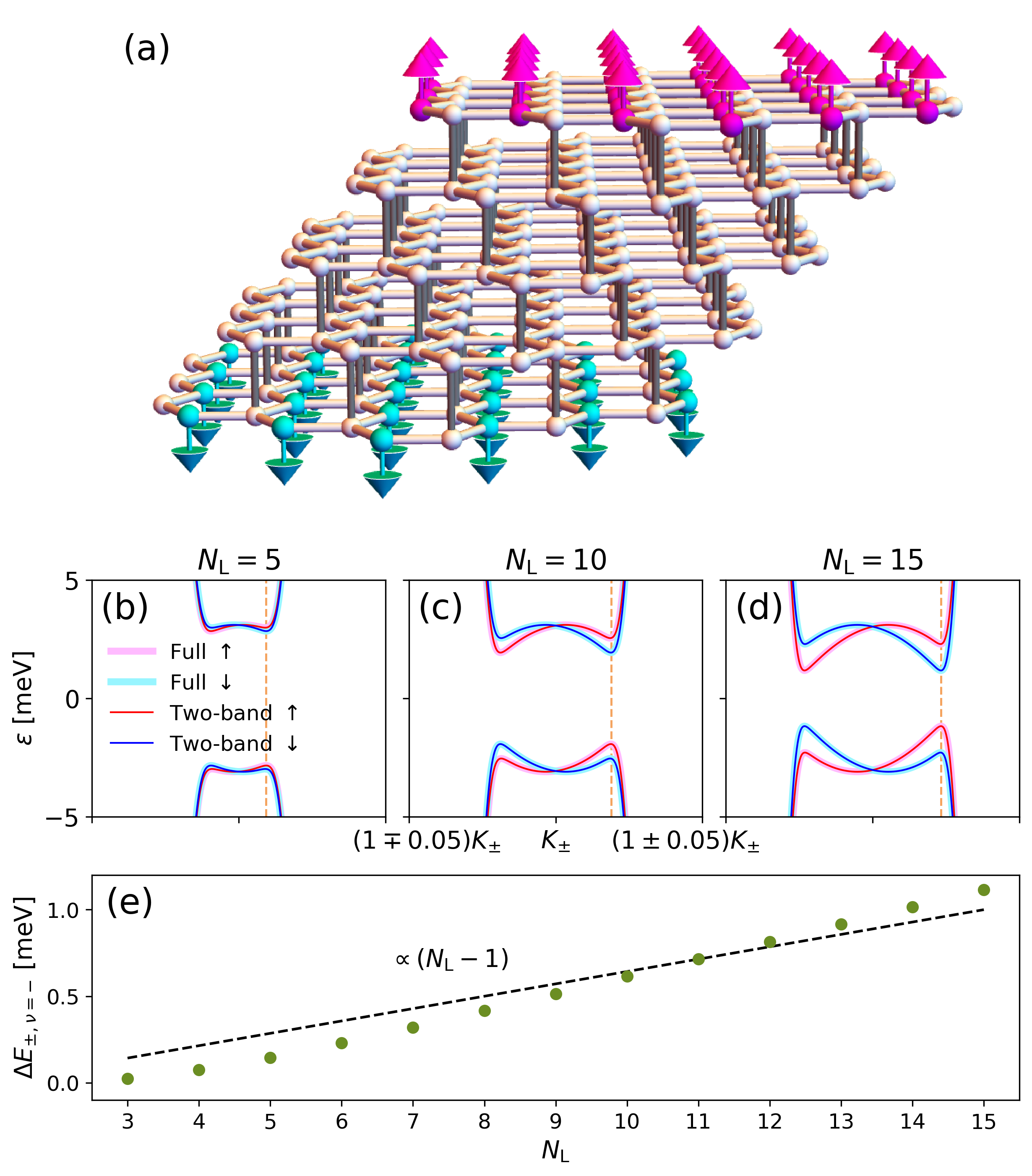}
    \caption{
        (a) Schematic of rhombohedral multilayer graphene with ACBACB$\cdots$ stacking. Magenta and cyan arrows indicate the opposite surface exchange fields associated with the layer AFM order.
        (b)--(d) Low-energy band dispersions obtained from the full tight-binding model (light colors) and the effective two-band model (dark colors) for $N_\mathrm{L}=5$, 10, and 15, respectively.
        The effective-model dispersions are shown only within the validity range $|c_0\kappa_{\pm,l}/t_1|<1$ for all layers $l$.
        (e) Layer-number dependence of the spin splitting $\Delta E$ at the Mexican-hat edges of the valence bands, which are indicated by orange dashed lines in (b)--(d). The dashed line indicates the linear dependence on $N_\mathrm{L}-1$ expected from the leading short-wavelength form factor. Parameters are $(t_0,t_1,m)=(-3100,+380,+6.2)$~[meV], $t_2=t_3=t_4=0$, and $(B_x,B_y)=(0,25.1)$~[T].
    }
    \label{Fig01}
\end{figure}

\begin{figure*}[t]
    \centering
    \includegraphics[width=0.95\textwidth]{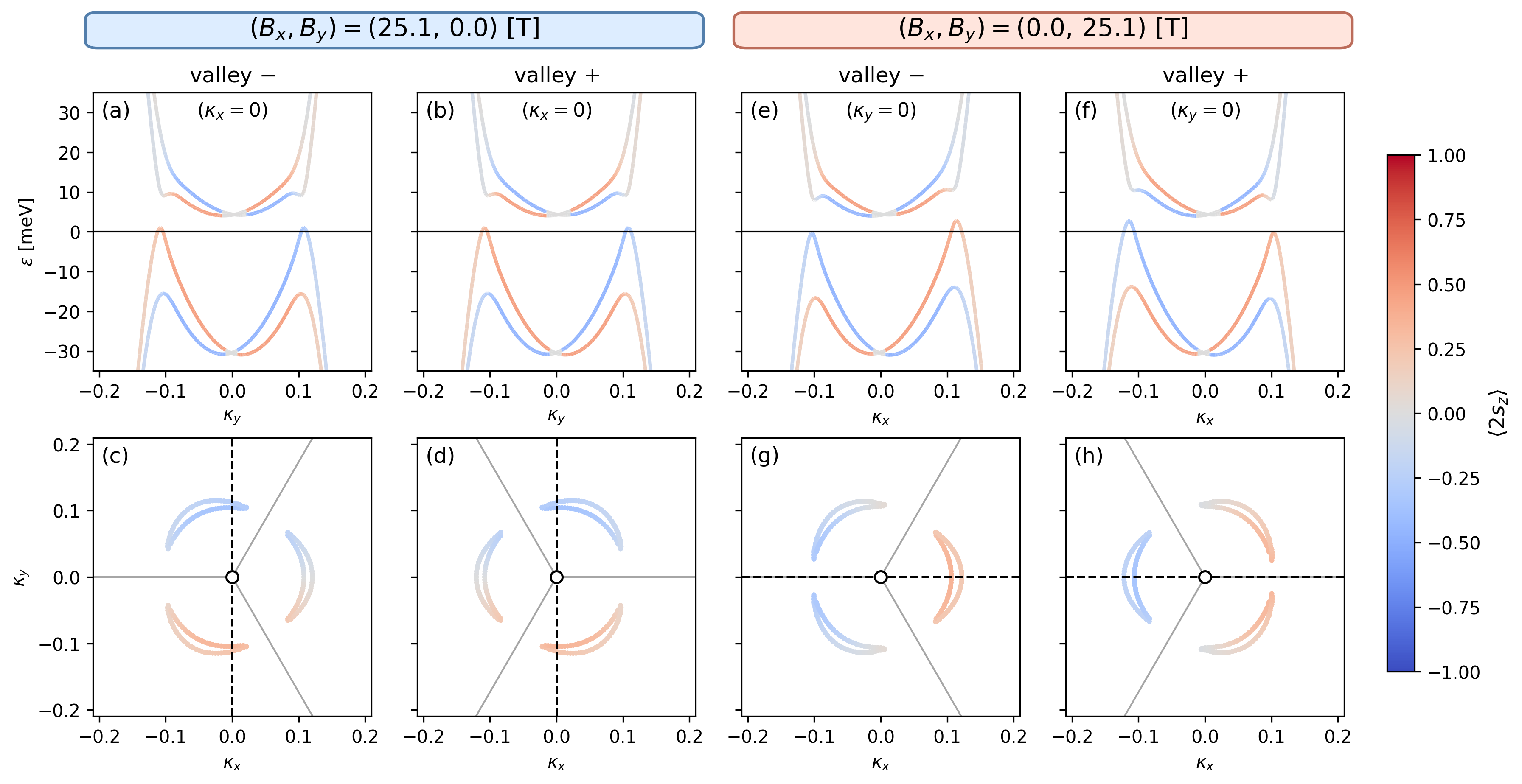}
    \caption{
        Self-consistent Hartree-Fock results for hole-doped RMG Hubbard model with $N_\mathrm{L}=10$ and $\Delta n=-(1/4N_\mathrm{L})\times2\cdot10^{-4}$ $(=-3.82\cdot10^{11}~[\mathrm{cm}^{-2}])$. 
        (a),(b) Valley-resolved band dispersions for an in-plane field along the $x$ direction, $(B_x,B_y)=(25.1,0)$~[T], plotted along $\kappa_x=0$. 
        (c),(d) Corresponding Fermi contours around the two valleys. (e),(f) Band dispersions for a field along the $y$ direction, $(B_x,B_y)=(0,25.1)$~[T], plotted along $\kappa_y=0$. 
        (g),(h) Corresponding Fermi contours. The color scale represents the spin polarization $\langle \sigma_z\rangle=\langle 2s_z\rangle$. Solid gray lines indicate the high-symmetry directions through each valley.
        Black dashed lines in lower panels denote momentum paths used in the upper panels.
    }
    \label{Fig02}
\end{figure*}

Rhombohedral multilayer graphene (RMG) has emerged as a remarkably rich platform for correlated-electron physics, owing to its low-energy flat bands~\cite{Guinea2006,Latil2006,HMin2008,Koshino2009} and their electrical tunability~\cite{CHLui2011,WBao2011,Yankowitz2013}. The low-energy states are predominantly localized on the two outermost layers of the rhombohedral stack, providing a distinctive surface-state platform for interaction-driven phenomena~\cite{Bernevig2025arXiv}. A wide range of correlated phases has been observed in RMG, including spontaneous symmetry-broken states~\cite{HZhou2021_halfmetal_TLG,ZGLu2024,Arp2024,WYLiao2025}, correlated insulators~\cite{YLee2014,GChen2019,KLiu2024,THan2024}, and superconductivity~\cite{HZhou2021_superconductivity_TLG,THan2025,JSeo2026}. This surface-state character also makes the layer number a natural control parameter for the low-energy electronic structure~\cite{YShi2020,YZhang2025,ZYChen2026,Lyu2026arXiv}. Within the minimal chiral description, the low-energy bands near each valley become progressively flatter with increasing layer number~\cite{HMin2008}, while experiments have revealed pronounced layer-dependent reconstruction of the electronic spectrum and band topology. One interesting magnetic phase of RMG is the ``layer antiferromagnet" with anti-aligned ferromagnetic top and bottom layers, which has been predicted theoretically early on~\cite{otani2010intrinsic,zhang2011spontaneous,DHXu2012,JJung2013} and subsequently observed experimentally~\cite{WBao2011,YLee2014,YFLiu2025}.

Meanwhile, inspired by the recent recognition of even-parity altermagnetism~\cite{LSmejkal2022,LSmejkal2022reviewPRX,IMazin2022editorialPRX}, nonrelativistic spin-split electronic states in magnetic materials have attracted significant attention. Comprehensive classifications based on spin-group symmetries~\cite{XJLuo2025arXiv,YTLiu2026,HYMa2026,ZYSong2026} have revealed that, beyond even-parity spin splitting in collinear altermagnets, odd-parity spin splitting can emerge in the nonrelativistic limit in noncollinear magnets~\cite{Hellenes2023arXiv,Brekke2024,YYu2025}. A growing number of candidate materials hosting such states, including insulating multiferroic~\cite{QSong2025} and metallic helimagnets~\cite{Yamada2025}, have subsequently been identified. Because these spin splittings do not rely on relativistic spin-orbit coupling, they can in principle reach energy scales far exceeding those generated by conventional relativistic mechanisms, making them attractive for spintronic applications.

Odd-parity spin splitting in a \textit{collinear} magnet, however, poses a fundamental challenge in the nonrelativistic limit. While even-parity altermagnetic spin splitting is compatible with conventional collinear magnetic order, odd-parity spin splitting is forbidden by the corresponding spin-group constraints unless an additional time-reversal-odd orbital degree of freedom is present~\cite{XJLuo2025arXiv}. Recent proposals circumvent this obstruction by invoking spontaneous orbital-current order~\cite{YPLin2026,Leeb2026}. 
Such realizations of collinear odd-parity magnetism, however, rely on additional electronic instabilities~\cite{Mielke2022,MChristensen2022,JZhan2026,Suetsugu2026} or demanding non-equilibrium driving protocols~\cite{SAAGhorashi2025,PHFu2026,SHuang2026,TZhu2026,BLi2026,TZhang2026,DLiu2026} and may therefore be less straightforward to engineer and prone to heating effects.

In this Letter, we demonstrate a simpler static route: an applied magnetic field can provide the required orbital time-reversal symmetry breaking without relying on relativistic spin-orbit coupling. In RMG, the orbital coupling to an in-plane magnetic field, combined with interaction-driven layer antiferromagnetism (AFM)~\cite{DHXu2012,JJung2013,YLee2014}, generates an odd-in-momentum spin splitting even when both spin-orbit and Zeeman couplings are absent. We identify the resulting state as a field-induced nonrelativistic collinear $p$-wave magnet. Since the spin splitting scales with the magnetic flux~\cite{Leeb2026} we find a linear dependence on the layer number, consistent with a low-energy two-band theory. We also show within self-consistent Hartree-Fock calculations that the resulting odd-wave magnetism is robust against realistic band-deformation effects, finite doping, and Zeeman coupling. Finally, we discuss distinct transport signatures in the form of a strong and tunable Edelstein effect.

\textit{Minimal Two-Band Model}.---To expose the mechanism of the $p$-wave spin splitting, we first consider a minimal tight-binding model for an $N_\mathrm{L}$-layer honeycomb lattice with ACBACB$\cdots$-type stacking in an external in-plane magnetic field $\mathbf{B}=(B_x,B_y,0)^\top$:
\begin{equation}
\begin{aligned}
    \label{eq:Hamiltonian}
    \mathcal{H}_{\mathrm{tb}}
    &= \sum_{\langle i,j \rangle} \sum_\sigma t_{ij} \left( e^{\mathrm{i}(e/\hbar)\mathbf{A}\cdot\mathbf{r}_{ij}} \hat{c}^\dagger_{i\sigma} \hat{c}_{j\sigma} + \mathrm{h.c.} \right) \\
    &\quad - \frac{m}{2} \sum_{\sigma\sigma'} \left( 
      A^\dagger_{l=1,\sigma}[\sigma^z]_{\sigma\sigma'}A_{l=1,\sigma'} 
    - B^\dagger_{l=N_\mathrm{L},\sigma}[\sigma^z]_{\sigma\sigma'}B_{l=N_\mathrm{L},\sigma'} 
    \right)
\end{aligned}
\end{equation}
The model is illustrated schematically in Fig.~\ref{Fig01}(a) and more precisely in Fig.~\ref{fig:tightbinding} in the End Matter. Here, $\hat{c}^\dagger_{i\sigma}$ $(=A^\dagger_{l\sigma}/B^\dagger_{l\sigma})$ creates an electron on sublattice A/B of layer $l$ with spin $\sigma(=\uparrow,\downarrow\mathrm{or}~\pm)$. The hopping amplitudes $t_0$, $t_1$, $t_2$, $t_3$, and $t_4$ enter the first term, with $\mathbf{r}_{ij}=\mathbf{r}_j-\mathbf{r}_i$. The in-plane field is incorporated through the Peierls phase using the Landau-gauge vector potential $\mathbf{A}=\left\{z-(N_\mathrm{L}+1)/2\right\}(B_y,-B_x,0)^\top$. The second term represents collinear layer AFM~\cite{DHXu2012,JJung2013,YLee2014}, with opposite exchange fields $m$ on the two surface sites, as indicated by the magenta and cyan arrows in Fig.~\ref{Fig01}(a). At this stage, we omit the Zeeman coupling to isolate the orbital-field-induced spin splitting.

Expanding the tight-binding Hamiltonian Eq.~(\ref{eq:Hamiltonian}) around the valleys $K_\pm=(\pm4\pi/3,0)$, we obtain
\begin{equation}
    \mathcal{H}_\mathrm{tb}(K_\xi+\bm{\kappa}) = 
    \Phi_{\xi\sigma}^\dagger(K_\xi+\bm{\kappa}) 
    H_{\xi\sigma}(K_\xi+\bm{\kappa})
    \Phi_{\xi\sigma}(K_\xi+\bm{\kappa}) 
\end{equation}
with
\begin{equation}
\begin{aligned}
    \label{eq:Htb_mat}
    &\quad\quad H_{\xi\sigma}(K_\xi+\bm{\kappa}) \\
    &= \left( \begin{array}{cccccccc}
        -\sigma m/2 & c_0 \kappa_{-,1} & & & & & & \\
        c_0 \kappa_{+,1} & & t_1 & & & & \\
        & t_1 & & c_0 \kappa_{-,2} & & & & \\
        & & c_0 \kappa_{+,2} & & t_1 & & & \\
        & & & t_1 & & \ddots & & \\
        & & & & \ddots & & t_1 & \\
        & & & & & t_1 & & c_0 \kappa_{-,N_\mathrm{L}} \\
        & & & & & & c_0 \kappa_{+,N_\mathrm{L}} & \sigma m/2
    \end{array} \right),
\end{aligned}
\end{equation}
where further-neighbor hopping $t_2$, $t_3$, and $t_4$ are omitted for simplicity and for maintaining chiral symmetry, and $c_0=-\sqrt{3}t_0/2$. $\xi(=\pm)$ labels the valleys and $\bm{\kappa}=(\kappa_x,\kappa_y)$ denotes the momentum measured from $K_\xi$. The basis is
\begin{equation}
\begin{aligned}
    \Phi^\dagger_{\xi\sigma}(K_\xi+\bm{\kappa}) &= \left( A^\dagger_{l=1,\sigma}(K_\xi+\bm{\kappa}),
    B^\dagger_{l=1,\sigma}(K_\xi+\bm{\kappa}),
    \cdots, \right. \\
    &\quad\quad\quad \left. \cdots,
    A^\dagger_{l=N_\mathrm{L},\sigma}(K_\xi+\bm{\kappa}),
    B^\dagger_{l=N_\mathrm{L},\sigma}(K_\xi+\bm{\kappa}) 
    \right),
\end{aligned}
\end{equation}
and the magnetic field enters through
\begin{equation}
    \kappa_{\pm,l} 
    = \xi \left\{ \kappa_x+\frac{ead}{\hbar}B_y \left(l-\frac{N_\mathrm{L}+1}{2} \right) \right\}
    \pm \mathrm{i} \left\{ \kappa_y-\frac{ead}{\hbar}B_x \left(l-\frac{N_\mathrm{L}+1}{2} \right) \right\}.
\end{equation}
Note that momenta are expressed in the dimensionless convention $\bm{\kappa}a$.
$a~(=0.246~[\mathrm{nm}])$ and $d~(=0.335~[\mathrm{nm}])$ are the intralayer lattice constant and the interlayer spacing of the RMG, respectively.
Crucially, the orbital coupling therefore shifts the momentum differently in each layer. This layer dependence distinguishes the two surface-localized low-energy states and, in the presence of their opposite exchange fields, provides the microscopic origin of the spin splitting.

We make this connection explicit by projecting onto the low-energy surface states through the Schrieffer-Wolff transformation
\begin{equation}
\begin{aligned}
    \label{eq:Schrieffer-Wolff}
    H^{\mathrm{eff}}_{\mathsf{P}} &= \mathsf{P}H\mathsf{P} + \mathsf{P}H\mathsf{Q}[E-H]^{-1}\mathsf{Q}H\mathsf{P} \\
    &= \mathsf{P}H\mathsf{P} 
    -  \mathsf{P}H\mathsf{Q}H^{-1}\mathsf{Q}H\mathsf{P}
    - E\mathsf{P}H\mathsf{Q}H^{-2}\mathsf{Q}H\mathsf{P} + \mathcal{O}(E^2)
\end{aligned}
\end{equation}
with
\begin{equation}
    \mathsf{P} = 
    \bigl| {A_{l=1}} \bigr\rangle 
    \bigl\langle A_{l=1} \bigr|
    \oplus
    \bigl|  B_{l=N_\mathrm{L}} \bigr\rangle
    \bigl\langle B_{l=N_\mathrm{L}} \bigr| , \ \ \
    \mathsf{Q} = \mathsf{1}-\mathsf{P}.
\end{equation}
Keeping terms up to first order in $E$ yields the \textit{generalized} eigenvalue problem
\begin{equation}
    \label{eq:generalized_eigenvalue_problem}
    H^{\mathrm{eff}}_{\xi\sigma}(\bm{\kappa}) \mathcal{U}_{\xi\sigma}(\bm{\kappa}) = E\left[ \mathsf{1}_{2\times2} + \mathfrak{G}(\bm{\kappa}) \right]\mathcal{U}_{\xi\sigma}(\bm{\kappa}).
\end{equation}
The effective Hamiltonian $H^{\mathrm{eff}}_{\xi\sigma}(\bm{\kappa})$ and the propagator $\mathfrak{G}(\bm{\kappa})$ are respectively given by
\begin{equation}
\begin{aligned}
    H^{\mathrm{eff}}_{\xi\sigma}(\bm{\kappa}) &= \left( \begin{array}{cc} 
        -\sigma m/2 & f^*(\bm{\kappa}) \\ f(\bm{\kappa}) & \sigma m/2
    \end{array} \right), \\
    \mathfrak{G}(\bm{\kappa}) &= \left( \begin{array}{cc} 
        \Sigma_1(\bm{\kappa}) & 0 \\ 0 & \Sigma_{N_\mathrm{L}}(\bm{\kappa})
    \end{array} \right),
\end{aligned}
\end{equation}
where
\begin{align}
    f(\bm{\kappa}) &= (-t_1)^{1-N_\mathrm{L}}\prod_{l=1}^{N_\mathrm{L}} \left(c_0\kappa_{+,l}\right), \\
    \Sigma_1(\bm{\kappa}) &= \sum_{r=1}^{N_\mathrm{L}-1} \left(\frac{c_0}{t_1}\right)^{2r} \prod_{l=1}^r |\kappa_{+,l}|^2, \\
    \Sigma_{N_\mathrm{L}}(\bm{\kappa}) &= \sum_{r=1}^{N_\mathrm{L}-1} \left(\frac{c_0}{t_1}\right)^{2r} \prod_{l=N_\mathrm{L}+1-r}^{N_\mathrm{L}} |\kappa_{+,l}|^2.
\end{align}
Note that the propagator $\mathfrak{G}(\bm{\kappa})$ is valley-independent. Importantly, $\Sigma_1(\bm{\kappa})$ and $\Sigma_{N_\mathrm{L}}(\bm{\kappa})$ encode the momentum-dependent dressing of the two surface states by the high-energy bulk degrees of freedom. The orbital field makes these two contributions inequivalent at a given $\bm{\kappa}$, while the opposite surface exchange fields convert this asymmetry into a spin-dependent energy shift.

Solving Eq.~(\ref{eq:generalized_eigenvalue_problem}) gives the energy spectrum $E_{{\xi\sigma},\nu}(\bm{\kappa})$, where $\nu=\pm$ is the band index. For the valence band, the resulting spin-splitting form factor near the valley is
\begin{equation}
\begin{aligned}
    \label{eq:form_factor}
    \Delta E_{\xi,\nu=-}(\bm{\kappa}) &= E_{\xi\uparrow,\nu=-}(\bm{\kappa}) - E_{\xi\downarrow,\nu=-}(\bm{\kappa}) \\
    &\propto \frac{m}{2}\left( \frac{1}{1+\Sigma_1(\bm{\kappa})} - \frac{1}{1+\Sigma_{N_\mathrm{L}}(\bm{\kappa})} \right) \\
    &\approx \frac{ead}{\hbar} \left(\frac{c_0}{t_1}\right)^2 m(N_\mathrm{L}-1) [\bm{\kappa}\times\mathbf{B}]_z,
\end{aligned}
\end{equation}
which directly establishes the $p$-wave character of the spin splitting. Its orientation is controlled by the in-plane magnetic field through $[\bm{\kappa}\times\mathbf{B}]_z$, while its leading magnitude grows as $N_\mathrm{L}-1$. This linear scaling has a simple orbital origin: the orbital magnetic flux associated with the separation between the two outermost layers is $(N_\mathrm{L}-1)$ times that associated with two adjacent layers, since their separation is $(N_\mathrm{L}-1)d$. The resulting orbital distinction between the two surface states therefore increases linearly with the multilayer thickness. We thus identify RMG in an external in-plane field as a highly tunable \textit{collinear} $p$-wave magnet.

To illustrate these results, we evaluate the low-energy dispersions for 5, 10, and 15 layers using realistic values of $(t_0,t_1,m)$, as shown in Figs.~\ref{Fig01}(b)--(d). The band dispersions obtained with the low-energy effective theory almost perfectly reproduce those obtained by straightforwardly diagonalizing Eq.~(\ref{eq:Htb_mat}), highlighting its validity. The characteristic Mexican-hat-like bands develop a pronounced spin splitting around their edges, which rapidly increases with layer number. Figure~\ref{Fig01}(e) summarizes this enhancement. While Eq.~(\ref{eq:form_factor}) gives a leading $N_\mathrm{L}-1$ dependence of the short-wavelength form factor, the spin splitting at the band edge exhibits an even stronger layer-number dependence.

\begin{figure}[!t]
    \centering
    \includegraphics[width=0.48\textwidth]{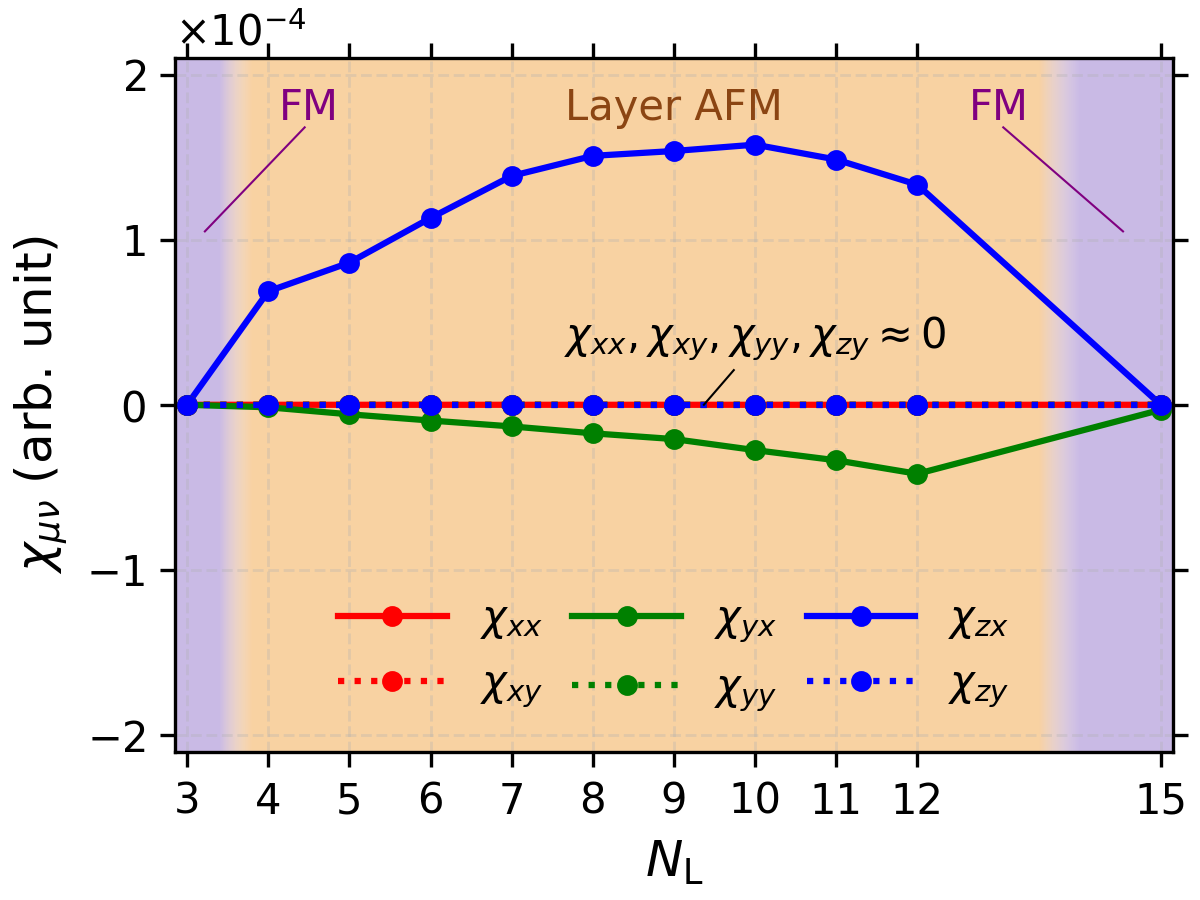}
    \caption{
        Layer-number dependence of the magnetic Edelstein susceptibility $\chi_{\mu\nu}$ for an in-plane field $(B_x,B_y)=(0,25.1)$~[T] and finite hole-doping of $\Delta n=-(1/4N_\mathrm{L})\times2\cdot10^{-4}$ $(=-3.82\cdot10^{11}~[\mathrm{cm}^{-2}])$. The shaded regions indicate the magnetic ground states obtained from the self-consistent Hartree-Fock calculation: field-polarized ferromagnetic (FM) order for small and large $N_\mathrm{L}$ and layer antiferromagnetic (AFM) order at intermediate $N_\mathrm{L}$.
        No converged solutions were obtained for \(N_L=13,14\), and these points are therefore omitted. 
    }
    \label{Fig03}
\end{figure}

\textit{Stability against Hole Doping}.---While the minimal chiral two-band model captures the essential mechanism of the $p$-wave spin splitting, its applicability to doped RMG requires further examination. In particular, remote hopping processes generate trigonal warping~\cite{Koshino2009} and substantially reconstruct the low-energy dispersion, including the van Hove singularities relevant to interaction-driven instabilities. Moreover, the layer AFM order was introduced phenomenologically through the exchange field $m$ in the two-band model. Upon hole doping, its stability is nontrivial because the enhanced density of states near the van Hove singularities can promote competing Stoner instabilities. We therefore examine whether the layer AFM order and the associated $p$-wave spin splitting survive in a realistic interacting model without assuming the magnetic order \textit{a priori}. To this end, we consider a Hubbard model including the full set of hopping parameters under an in-plane magnetic field, whose Hamiltonian is given by
\begin{equation}
\begin{aligned}
    \label{eq:Hubbard}
    \mathcal{H}_\mathrm{Hubbard} 
    &= \sum_{\langle i,j \rangle} \sum_\sigma t_{ij} \left( e^{\mathrm{i}(e/\hbar)\mathbf{A}\cdot\mathbf{r}_{ij}} \hat{c}^\dagger_{i\sigma} \hat{c}_{j\sigma} + \mathrm{h.c.} \right)
    + U\sum_i \hat{n}_{i\uparrow} \hat{n}_{i\downarrow} \\
    &\quad\quad + \frac{g\mu_\mathrm{B}}{2} \mathbf{B}\cdot \sum_i \sum_{\sigma\sigma'} \hat{c}^\dagger_{i\sigma}[\bm{\sigma}_i]_{\sigma\sigma'} \hat{c}_{i\sigma'}.
\end{aligned}
\end{equation}
The first term describes electron hopping and is identical to that in Eq.~(\ref{eq:Hamiltonian}). The second term represents the onsite repulsive Coulomb interaction with interaction strength $U$, where $\hat{n}_{i\sigma}=\hat{c}^\dagger_{i\sigma}\hat{c}_{i\sigma}$. The third term accounts for the Zeeman coupling, where $\mu_\mathrm{B}$ is the Bohr magneton and the $g$ factor is set to $+2$. We use $(t_0,t_1,2t_2,t_3,t_4)=(-3100,+380,-21,+290,+141)$~[meV]~\cite{AZibrov2018,Uchida2026} and $U=+6200$~[meV]. We deliberately choose a relatively conservative interaction strength, noting that the effective onsite repulsion can be larger depending on the degree of screening~\cite{Wehling2011,Schueler2013,Roesner2015}. This choice allows us to assess the stability of the layer AFM order upon hole doping without relying on a particularly strong interaction.

We solve Eq.~(\ref{eq:Hubbard}) directly on the full lattice within the self-consistent Hartree-Fock approximation by decomposing the interaction term as
\begin{equation}
\begin{aligned}
    \hat{n}_{i\uparrow} \hat{n}_{i\downarrow} &\approx 
      \langle \hat{n}_{i\uparrow} \rangle \hat{n}_{i\downarrow}
    + \hat{n}_{i\uparrow} \langle \hat{n}_{i\downarrow} \rangle
    - \langle \hat{n}_{i\uparrow} \rangle \langle \hat{n}_{i\downarrow} \rangle
    \\ &\quad
    - \langle \hat{c}^\dagger_{i\uparrow}\hat{c}_{i\downarrow} \rangle \hat{c}^\dagger_{i\downarrow}\hat{c}_{i\uparrow} 
    - \hat{c}^\dagger_{i\uparrow}\hat{c}_{i\downarrow} \langle \hat{c}^\dagger_{i\downarrow}\hat{c}_{i\uparrow} \rangle
    + \langle \hat{c}^\dagger_{i\uparrow}\hat{c}_{i\downarrow} \rangle \langle \hat{c}^\dagger_{i\downarrow}\hat{c}_{i\uparrow} \rangle,
\end{aligned}
\end{equation}
with no valley expansion or low-energy projection and no imposed magnetic order. Without loss of generality we fix the orientation of the N\'{e}el vector along $z$-direction in the layer AFM solutions. The field strength considered here, 25.1~T, is experimentally accessible, with in-plane fields exceeding 30~T having already been employed in multilayer graphene devices~\cite{YLee2014}. Figure~\ref{Fig02} shows the resulting low-energy electronic structure for $N_\mathrm{L}=10$ at finite hole doping of $\Delta n=-(1/4N_\mathrm{L})\times2\cdot10^{-4}$ $(=-3.82\cdot10^{11}~[\mathrm{cm}^{-2}])$, where $n=0.5$ corresponds to charge neutrality, while $1/(4N_L)$ corresponds to the filling of a single spin-resolved band over the full Brillouin zone. Despite the substantial reconstruction of the low-energy bands and Fermi surfaces by the remote hopping processes, a layer AFM solution remains stable and exhibits a clear momentum-dependent spin splitting near the Fermi level. For $\mathbf{B}\parallel\hat{x}$ [Figs.~\ref{Fig02}(a)--(d)], the spin splitting is most pronounced along the momentum direction perpendicular to the field, while rotating the field to $\mathbf{B}\parallel\hat{y}$ [Figs.~\ref{Fig02}(e)--(h)] correspondingly rotates the characteristic anisotropy of the spin splitting. This behavior is consistent with the $[\bm{\kappa}\times\mathbf{B}]_z$ dependence obtained from the minimal chiral two-band model. The maximal spin splitting reaches nearly 20~meV. These results confirm that the characteristic $p$-wave spin splitting persists in the self-consistent layer AFM state even in the presence of realistic band-structure effects, finite hole doping, and Zeeman coupling.

\textit{Magnetic Edelstein Effect}.---We finally discuss a transport signature of the field-induced $p$-wave spin splitting~\cite{Hellenes2023arXiv,Pari2025,Chakraborty2025,Habel2026arXiv}.
It has been pointed out that collinear $p$-wave magnets exhibit finite Edelstein response when a Zeeman coupling perpendicular to the N\'{e}el vector exists~\cite{Leeb2026}, which is referred to as the \textit{magnetic} Edelstein effect.
Within the semiclassical Boltzmann formalism, an applied electric field $\mathbf{E}$ produces a nonequilibrium spin polarization $\delta s_\mu=\chi_{\mu\nu}E_\nu$ $(\mu=x,y,z~\mathrm{and}~\nu=x,y)$, where, in the constant-relaxation-time approximation,
\begin{equation}
    \chi_{\mu\nu}
    \approx
    e\hbar\tau
    \sum_n
    \int_{\mathrm{BZ}}
    \frac{d^2k}{(2\pi)^2}
    s_{\mu,\mathbf{k}n}
    v_{\nu,\mathbf{k}n}
    \left(
    -\frac{\partial f}{\partial E_{\mathbf{k}n}}
    \right).
\end{equation}
Here, $s_{\mu,\mathbf{k}n}$ and $v_{\nu,\mathbf{k}n}$ are the spin expectation value and group velocity of band $n$, respectively, $f$ is the Fermi distribution function, and $\tau$ is a momentum- and band-independent relaxation time. This treatment retains the intraband contribution to the current-induced spin polarization and neglects interband-coherence and vertex-correction effects.

Figure~\ref{Fig03} shows the layer-number dependence of $\chi_{\mu\nu}$ calculated from the self-consistent Hartree-Fock band dispersions. In the field-polarized FM regimes at small and large $N_\mathrm{L}$, the Edelstein response is nearly absent. By contrast, finite components develop prominently within the surface-AFM regime. Most notably, $\chi_{zx}$ is by far the dominant contribution, reflecting the odd-in-momentum $z$-spin polarization generated by the $p_x$-wave spin splitting. Its magnitude substantially exceeds that of $\chi_{yx}$, which is associated with the field-induced spin polarization along the magnetic-field direction. The dominance of $\chi_{zx}$ therefore distinguishes the orbital-field-induced $p_x$-wave response from the conventional spin polarization induced directly by the Zeeman coupling. We note that the measured hierarchy depends on the layer sensitivity of the experimental probe. For a representative exponentially decaying sensitivity profile,  $\chi_{zx}$ is already dominant when the bottom-layer sensitivity remains about half that of the top layer [See End Matter]. Moreover, even if the symmetry between the top and bottom layer is broken explicitly the p-wave character can easily be probed by inverting the in-plane field direction, which reverses the orbital field induced magnetic moment. Overall, the pronounced magnetic Edelstein response confined to the surface-AFM regime provides a clear transport signature of the $p$-wave magnetic state.

\textit{Discussion}.---Our results establish static orbital coupling to a magnetic field as a new route to odd-parity spin splitting in collinear magnets. Unlike mechanisms based on spontaneous orbital-current order or nonequilibrium driving, the required time-reversal-odd orbital degree of freedom is supplied directly by the applied field. RMG provides a particularly favorable realization of this principle, as its surface-localized low-energy states convert the layer-dependent orbital coupling to an in-plane magnetic field into a strongly enhanced spin splitting. In this setting, the magnetic field does not merely tune an existing spin-split state, but generates and controls the momentum-space distribution of the $p$-wave spin splitting. As the splitting is set by the magnetic flux between the top and bottom layer sizeable splittings appear for moderate field strength which increase with increasing layer number.

Our perspective extends the design space of nonrelativistic spin-split states beyond magnetic order alone, by promoting orbital coupling to an external field to an active ingredient for engineering their spin-crystallographic symmetry. The magnitude of the $p$-wave spin splitting could be further enhanced through magnetic proximity engineering. For example, encapsulating RMG between out-of-plane ferromagnetic insulators with antiparallel magnetizations could induce opposite exchange fields on the two surfaces, reinforcing the intrinsic layer AFM order and thereby enhancing the $p$-wave spin splitting.

RMG further hosts a broad landscape of interaction-driven phases, raising the prospect of combining them with this externally controllable odd-parity spin-split electronic structure. Superconductivity~\cite{HZhou2021_superconductivity_TLG,THan2025,LDLNir2026arXiv,JSeo2026} is particularly intriguing in this regard: the magnitude and orientation of the normal-state $p$-wave spin splitting can be tuned independently through the layer number and the direction of the in-plane magnetic field. This provides a setting in which the influence of an odd-parity spin-momentum structure on superconducting pairing can be manipulated without relying on intrinsic spin-orbit coupling. More broadly, the combination of strong correlations, layer engineering, and orbital-field control may offer a versatile route to designing tunable spin-momentum structures in stacked two-dimensional quantum materials.

\textit{Note added}.---Upon completion of this work, we became aware of Ref.~\cite{ZTSun2026arXiv}, which independently discusses a related mechanism of orbital-field-induced spin splitting in compensated magnetic systems, in the broader context of probing antialtermagnetism.

\textit{Acknowledgment}.---We acknowledge helpful discussions with V. Leeb and J. Habel. R.E. acknowledges financial support by JSPS Overseas Fellowship. J.K. acknowledges support from the Deutsche Forschungsgemeinschaft (DFG, German Research Foundation) under grants TRR 360 - 492547816, KN1254/1-2, KN1254/2-1 and under Germany’s Excellence Strategy EXC-2111-390814868. J.K. acknowledges support from the Munich Quantum Valley, which is supported by the Bavarian state government with funds from the High-tech Agenda Bayern Plus. J.K. further acknowledges support from the Imperial-TUM flagship partnership and the Keck foundation.

\clearpage

\section{End Matter}

\subsection{Tight-binding model}

\begin{figure}[hbt]
    \centering
    \includegraphics[width=0.25\textwidth]{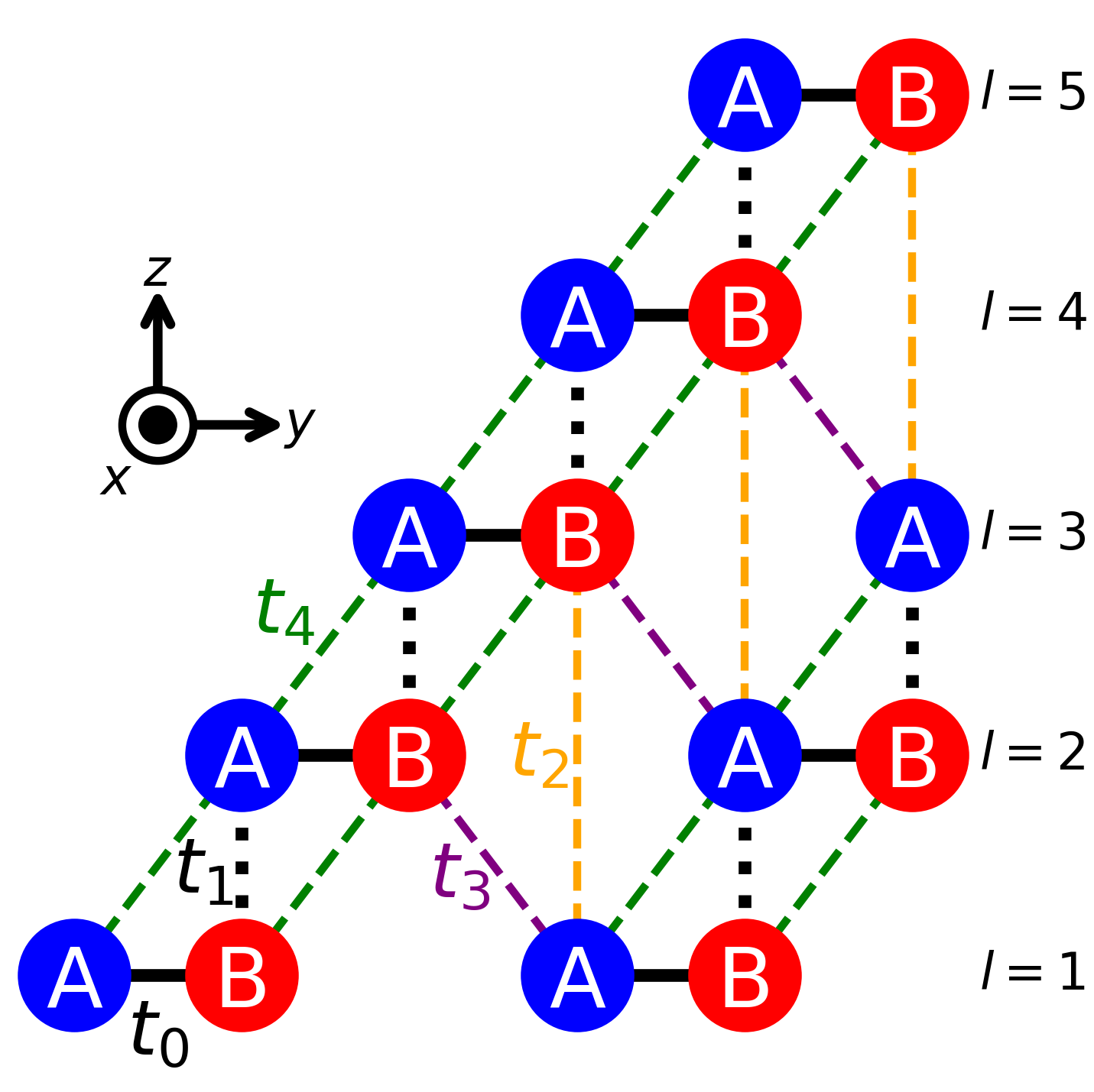}
    \caption{
        Schematic of the tight-binding model of the RMG.
    }
    \label{fig:tightbinding}
\end{figure}
As an example, we present the details of the tight-binding model of the rhombohedral pentalayer graphene in Fig.~\ref{fig:tightbinding}, in which the hopping integrals, $t_0$, $t_1$, $t_2$, $t_3$, and $t_4$, are assigned.

\subsection{Probe-depth dependence of the magnetic Edelstein effect}

\begin{figure}[hbt]
    \centering
    \includegraphics[width=0.48\textwidth]{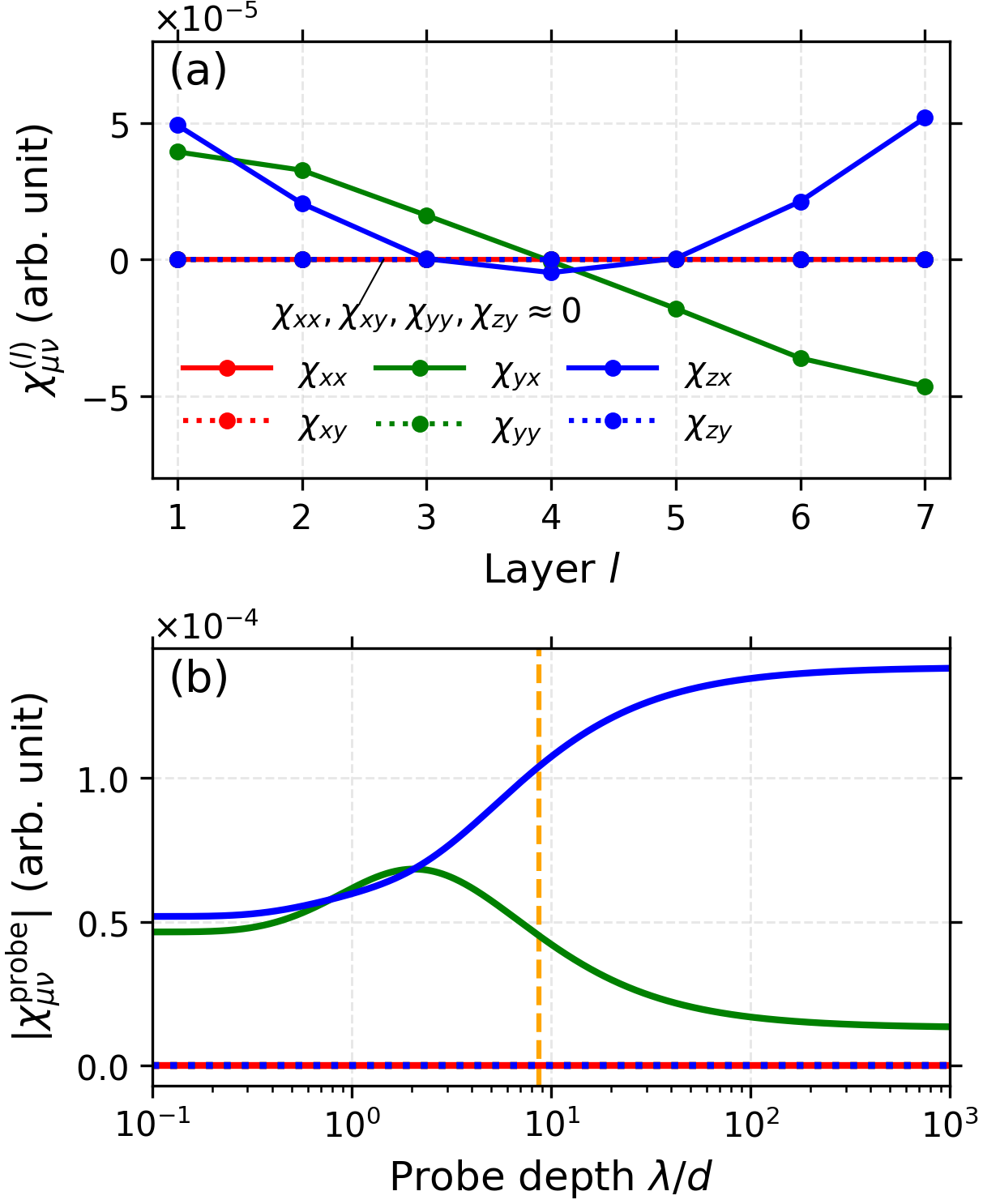}
    \caption{
        (a) Layer dependence and (b) probe-depth dependence of the magnetic Edelstein effect for \(N_{\mathrm L}=7\), $(B_x,B_y)=(0,25.1)$~[T], and $\Delta n=-(1/4N_\mathrm{L})\times2\cdot10^{-4}$ $(=-3.82\cdot10^{11}~[\mathrm{cm}^{-2}])$.
        In (b), the layer sensitivity is assumed to decay exponentially from the top layer, and the orange dashed line indicates the probe depth at which the sensitivity to the bottom layer is half that to the top layer.
    }
    \label{fig:edelstein_layer}
\end{figure}

To examine the effect of the layer sensitivity of an experimental probe, we decompose the magnetic Edelstein susceptibility into layer-resolved contributions. We define
\begin{equation}
    \chi_{\mu\nu}^{(l)}
    =
    e\hbar\tau
    \sum_n
    \int_{\mathrm{BZ}}
    \frac{d^2k}{(2\pi)^2}
    s_{\mu,\mathbf{k}n}^{(l)}
    v_{\nu,\mathbf{k}n}
    \left(
    -\frac{\partial f}{\partial E_{\mathbf{k}n}}
    \right),
\end{equation}
where
\begin{equation}
    s_{\mu,\mathbf{k}n}^{(l)}
    =
    \langle \mathbf{k}n |
    P_l \hat{s}_{\mu}
    | \mathbf{k}n \rangle ,
\end{equation}
and $P_l$ projects onto layer $l$. Since $\sum_l P_l=1$, the susceptibility used in the main text is recovered exactly by summing over all layers,
\begin{equation}
    \chi_{\mu\nu}
    =
    \sum_{l=1}^{N_{\mathrm L}}
    \chi_{\mu\nu}^{(l)}.
\end{equation}

Figure~\ref{fig:edelstein_layer}(a) shows the resulting layer-resolved response. The $\chi_{yx}^{(l)}$ contributions change sign across the multilayer and therefore substantially cancel in the layer-summed response. In contrast, the dominant surface contributions to $\chi_{zx}^{(l)}$ have the same sign and add predominantly constructively. This explains the dominance of $\chi_{zx}$ in the total magnetic Edelstein response shown in the main text.

To model the finite probe depth of a surface-sensitive measurement, we introduce an exponentially decaying layer sensitivity,
\begin{equation}
    w_l(\lambda)
    =
    \exp\left[
    -\frac{(N_{\mathrm L}-l)d}{\lambda}
    \right],
\end{equation}
where $l=N_{\mathrm L}$ denotes the exposed top layer, $d~(=0.335~[\mathrm{nm}])$ is the interlayer spacing, and $\lambda$ is the probe depth. The corresponding probe-weighted susceptibility is defined as
\begin{equation}
    \chi_{\mu\nu}^{\mathrm{probe}}(\lambda)
    =
    \sum_{l=1}^{N_{\mathrm L}}
    w_l(\lambda)\chi_{\mu\nu}^{(l)}.
\end{equation}
We do not normalize the weights, so that
$\chi_{\mu\nu}^{\mathrm{probe}}\to\chi_{\mu\nu}^{(N_{\mathrm L})}$
for $\lambda/d\to0$ and
$\chi_{\mu\nu}^{\mathrm{probe}}\to\chi_{\mu\nu}$
for $\lambda/d\to\infty$.

Figure~\ref{fig:edelstein_layer}(b) shows $|\chi_{\mu\nu}^{\mathrm{probe}}|$ as a function of $\lambda/d$. As the probe depth increases, the cancellation between the layer-resolved $\chi_{yx}^{(l)}$ contributions suppresses $|\chi_{yx}^{\mathrm{probe}}|$, whereas $|\chi_{zx}^{\mathrm{probe}}|$ approaches the dominant layer-summed response. The orange dashed line marks the probe depth at which the bottom-layer sensitivity is half that of the top layer,
\begin{equation}
    \frac{w_1}{w_{N_{\mathrm L}}}
    =
    \exp\left[
    -\frac{(N_{\mathrm L}-1)d}{\lambda}
    \right]
    =
    \frac{1}{2}.
\end{equation}
Already at this probe depth, $\chi_{zx}$ is clearly the dominant response component.

The probe-depth dependence considered here may also be relevant for optical detection of the magnetic Edelstein effect. In particular, the dominant $\chi_{zx}$ response corresponds to a current-induced out-of-plane spin polarization and could therefore, in principle, be detected through magneto-optical Kerr rotation. Graphene exhibits an optical opacity of only approximately 2.3\% for a monolayer in the visible range, with the opacity observed to increase approximately linearly with layer number~\cite{Nair2008}. This suggests that optical probes can retain appreciable sensitivity across the thickness of the RMG samples considered here, rather than being restricted to the exposed surface layer. Magneto-optical measurements may therefore provide a promising route to detecting the magnetic Edelstein response. A quantitative prediction of the Kerr signal, however, would require an explicit treatment of the dynamical magneto-optical response and is beyond the scope of the present work.

\subsection{Top-layer-projected Fermi surface}

\begin{figure}[hbt]
    \centering
    \includegraphics[width=0.50\textwidth]{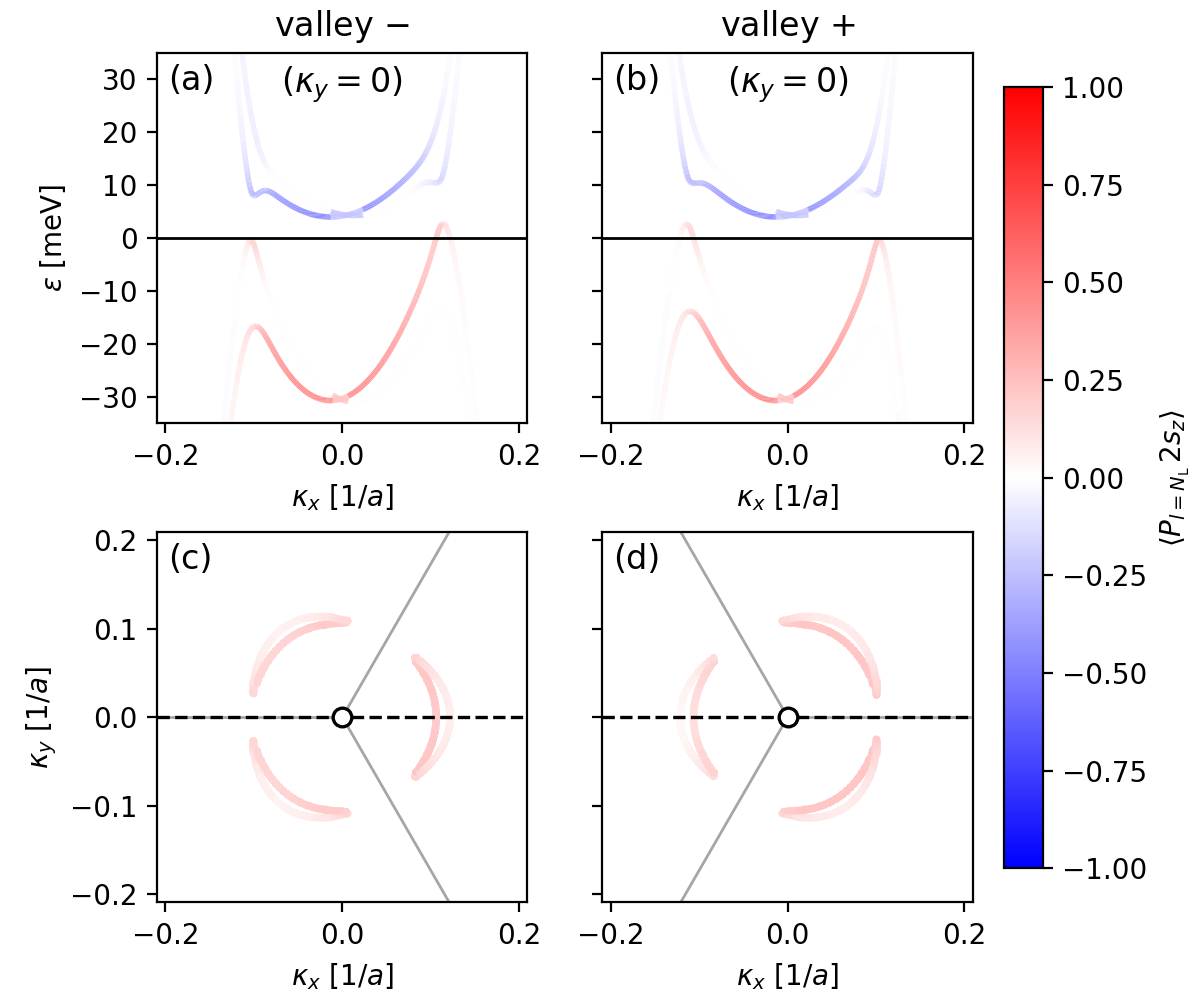}
    \caption{
        Top-layer-projected (a)(b) band dispersions and (c)(d) Fermi surfaces for $N_\mathrm{L}=10$, $(B_x,B_y)=(0,25.1)~[\mathrm{T}]$, and $\Delta n=-(1/4N_\mathrm{L})\times2\cdot10^{-4}$ $(=-3.82\cdot10^{11}~[\mathrm{cm}^{-2}])$.
    }
    \label{fig:edelstein_layer}
\end{figure}

Figure~\ref{fig:edelstein_layer} shows the band structure and Fermi surface projected onto the exposed top layer $(l=N_\mathrm{L})$. Because the low-energy states of RMG are predominantly localized on the two outermost layers and the layer-AFM order produces opposite spin polarizations on the two surfaces, the highest occupied states with appreciable top-layer weight are predominantly spin-up polarized. Consequently, the corresponding isoenergy contours on the top layer exhibit an approximately $s$-wave-like spin splitting, rather than directly displaying the odd-in-momentum spin texture of the full multilayer electronic structure. This observation has an important implication for surface-sensitive spectroscopic probes. Quantum twisting microscope~\cite{Inbar2023,JXiao2026,Pichler2024,NWei2025,Waschitz2026,NWei2026,MLee2026,Birkbeck2025,JXiao2024} and quasiparticle interference measurements by scanning tunneling microscopy~\cite{Avraham2018,Rhodes2026arXiv_QPI,YFLiu2025,Sakai2026} predominantly probe tunneling involving the exposed top layer and are therefore expected to be primarily sensitive to this surface-projected, effectively $s$-wave-like spin splitting. A direct spectroscopic identification of the $p$-wave character, which originates from the momentum-dependent relation between the oppositely spin-polarized top- and bottom-surface states, would instead require appreciable sensitivity to both outermost layers. These strongly surface-selective probes are therefore not ideally suited for a direct detection of the $p$-wave spin splitting, although they can still provide complementary information on the spin-polarized surface electronic structure.

\bibliography{bib}

\end{document}